\documentstyle[amssymb,aps,prl,epsfig,floats]{revtex}

\DeclareMathAlphabet{\mabold}{OT1}{cmr}{bx}{it}

\newcommand{\br}{{\mabold{r}}}
\newcommand{\bs}{{\mabold{s}}}

\newcommand{\bq}{{\mabold{q}}}
\newcommand{\bn}{{\mabold{n}}}

\newcommand{\bk}{{k}}
\newcommand{\bp}{{\mabold{p}}}
\newcommand{\nablab}{{\mabold{\nabla}}}

\newcommand{\Ocal}{{\cal O}}
\newcommand{\beq}{\begin{equation}}
\newcommand{\eeq}{\end{equation}}
\newcommand{\Hcal}{{\cal H}}
\newcommand{\Fcal}{{\cal F}}
\newcommand{\Pcal}{{\cal P}}
\newcommand\mean[1]{{\langle#1\rangle}}
\newcommand{\rh}{{\mathrm{h}}}
\newcommand{\rp}{{\mathrm{p}}}

\begin{document}


\draft
\twocolumn[\hsize\textwidth\columnwidth\hsize\csname @twocolumnfalse\endcsname

\title{A microscopic model for thin film spreading}
\author{
Douglas B.\ Abraham$^1$, Rodolfo Cuerno$^2$, and Esteban Moro$^1$}
\address{$^1$Theoretical Physics, University of Oxford, 1
Keble Road OX1 3NP, United Kingdom \\
$^2$Dpto.\ de Matem\'aticas and GISC,
Universidad Carlos III de Madrid, Avda.\ Universidad 30, 28911 Legan\'es,
Spain}
\date{\today}

\maketitle

\begin{abstract}
A microscopic, driven lattice gas model is proposed for the dynamics
and spatio-temporal fluctuations of the precursor film observed in
spreading experiments. Matter is transported both by holes and
particles, and the distribution of each can be described by driven
diffusion with a moving boundary. This picture leads to a stochastic
partial differential equation for the shape of the boundary, which
agrees with the simulations of the lattice gas. Preliminary results
for flow in a thermal gradient are discussed.
\end{abstract}

\pacs{PACS numbers: 68.08.Bc, 68.15+e, 64.60.Ht, 05.70.Np}

%

]

Spreading of involatile liquid drops on surfaces and fibers plays a
significant role in many technologies, the efficient deployment of
which requires detailed understanding of the underlying process
\cite{ball,lhf}. Behavior at a macroscopic scale is accurately
described by hydrodynamics \cite{deGennes}, but experiments performed
over the last 15 years or so have produced a surprise: in the case of
complete wetting, the spreading drop, when examined on an atomic
length scale in the direction normal to the substrate, is found to be
preceded by a precursor film about one molecule thick; this can be
followed laterally out to an extension of the order of $10^7$
molecules diameter. At the horizontal resolving power of the technique
used (ellipsometry), the precursor film is found to be flat and
homogeneous. Its radius advances with time as $\sqrt{t}$. Typical
systems showing this class of behavior which can be investigated by
ellipsometry are various silanes spreading over atomically flat
Si(111) wafers with highly pure oxydised surfaces. With selected
silanes, such systems even show dynamical layering with up to four
superposed precursor films advancing as $\sqrt{t}$, the layer directly
in contact with the substrate being much faster than the others.  As
Ball observed over a decade ago \cite{ball}, at that time there was
hardly even a formative theory of such phenomena. Such theory must
capture the experimentally determined diffusive
behavior and, at the same time, explain how an extremely viscous,
involatile material is transported from the reservoir to the precursor
edge.

A partial solution was first achieved by applying Molecular
Dynamics to a droplet composed of spherical molecules with
Lennard-Jones interactions \cite{md_molec}.  The strength of these is
adjusted to achieve sufficient involatility.  Experimentally, the
droplet is placed in contact with a substrate composed of much smaller
units; this is therefore treated as a continuum for calculating the
interaction of the spreading molecules with the substrate. Such a
model shows a precursor film with the correct, diffusive behavior
\cite{md_molec,md_chain}. Monte Carlo (MC) lattice gas simulations in 3d
\cite{douglas_ising,cazabat} confirm this, and suggest a dual mechanism of 
matter transport involving on the one hand {\em particles} located in
a supernatant layer directly placed above the precursor film, and on
the other, {\em holes} in the precursor film itself. At the same time,
essentially all the configurational change occurs at the boundary of
the drop. However, so far these ideas just constitute a picture
without quantitative implications. 

In this Letter, we introduce an associated 2d driven Ising lattice gas
model which can be simulated far more effectively. Extensive
continuous time MC simulations are consistent with the $\sqrt{t}$
behavior found in all the experiments to which we refer \cite{lhf}.
They also allow us to formulate a probabilistic continuum model of
independent hole and particle diffusion in the supernatant and
precursor films, which accounts for the matter transfer between the
reservoir and the precursor edge (PE). The motion of the latter in the
driven lattice gas is then equivalent to a moving boundary in the
continuum model. Starting from the constitutive equations of the
moving boundary problem, we derive the effective dynamics of the PE as
a stochastic partial differential equation. We thus obtain analytical
predictions which are consistent both with experiments and with the
original 2d Ising model, and which can be extended to the asymptotic
regime thus far unexplored both theoretically and experimentally.

We now describe our MC lattice gas simulations for this problem. At
each cell on a square lattice $\br =(x,y,z)$ with unit lattice
spacing, we define the ocuppation number $n(\br,t)$ to be $0$ or $1$
depending whether the cell is occupied or not. Moreover, we restrict
$z$ to the values $z=1$ (precursor film) and $z=2$ (supernatant
layer).  The interaction energy for a given configuration is
\cite{douglas_ising}
\beq\label{energy}
\Hcal = -J \sum_{|\br-\bs|=1} n(\br,t) n(\bs,t)
-A \sum_{z=1,2} \frac{1}{z^3} n(\br,t),
\eeq
where the sums are over $\br, \bs$ on the lattice. The last term is
the van der Waals interaction with the {\em continuum} substrate
characterized by a Hamaker constant $A > 0$. We are interested in
$J/k_B T$ large enough to achieve a high degree of involatility and
$A/k_B T$ large enough to be in the complete wetting regime
\cite{ref_wetting}. Continuous time MC Kawasaki dynamics are used 
\cite{MC_rate} with initial conditions: {\em (i)} $n(\br,0) = 1$
for $y=1$, but with any $x$ and for both $z=1$ and $z=2$; {\em (ii)}
$n(\br,0) = 0$ for $y > 1$. If, as a result of an allowed MC move at
any later stage, $n(x,1,z,t)=0$, then the implied vacancy is filled
immediately from the putative reservoir droplet, $n(x,1,z,t)=1$
being maintained in this sense for all $t > 0 $. Thus, the detailed
evolution of the reservoir droplet is ignored. In reality, it would
equilibrate relatively rapidly in profile by motion of particles near
its surface and deflate by emission of supernatant particles and
absorption of precursor holes \cite{deGennes}. 
The main advantage of the boundary
condition we use, and a very significant one, is in computability. We
believe our model to be new --- it is two superposed Ising lattice
gases driven from one edge.

\begin{figure}
\begin{center}
\epsfig{file=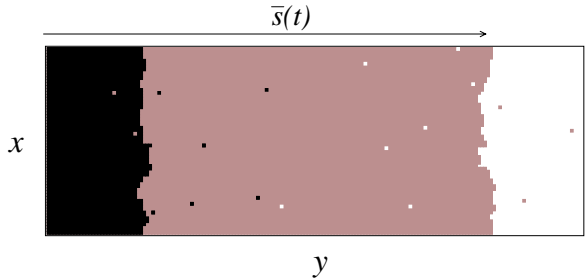, width=1.6in, angle=-90}
\caption{Top view of a typical snapshot of our lattice gas model for $t = 2
\times 10^7$ MC units. Occupied
cells in $z=1$ (precursor film) are in gray, while occupied
cells in $z=2$ (supernatant film) appear in black. Non-colored cells
are empty. Parameters used are $A=10$, $J=1$, $k_B T = 1/3$, $L_x=128$.}
\label{cartoon}
\vspace{-0.5cm}
\end{center}
\end{figure}
A typical snapshot of the simulations is shown in Fig.\ \ref{cartoon}.
There is a compact film advancing in both the $z=1$ and $z=2$ layers, 
but the former (shown black) advances much faster that the latter 
(shown gray). It is easy to understand why this should be so: if  
a hole in the precursor film diffuses to lie beneath the compact film
in $z=2$, at an appropriately selected update it will be filled by 
creating a hole in film above, which diffuses and may be trapped at 
the edge $y=1$, or at the supernatant film boundary, causing this 
film to shrink. The precursor advances because of emission of holes 
into itself, but also from the arrival of supernatant particles at 
its edge; such particles drop into the $z=1$ plane with concomitant
Hamaker stabilization \cite{no_freespace}. 

\begin{figure}
\begin{center}
\epsfig{file=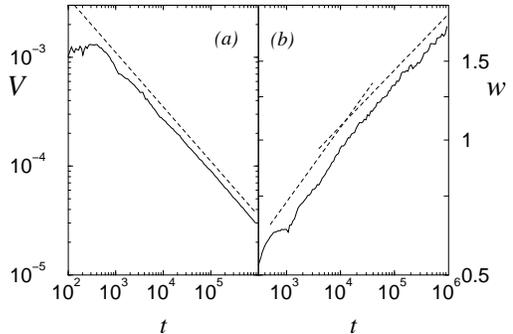, width=2.1in, angle=-90}
\caption{Log-log plot of the instantaneous velocity [panel (a)] and roughness
[panel (b)] of the PE. Solid lines are averages over 200 Monte
Carlo realizations with $A=10$, $J=1$, $k_B T = 1/3$, and $L_x=256$. All units
are arbitrary. Dashed line in panel (a) depicts the $t^{-1/2}$ behavior, 
while those in panel (b) correspond to the $t^{1/6}$ and $t^{1/8}$ scaling 
behaviors.}\label{w2}
\vspace{-0.5cm}
\end{center}
\end{figure}

\begin{figure}[t]
\begin{center}
\epsfig{file=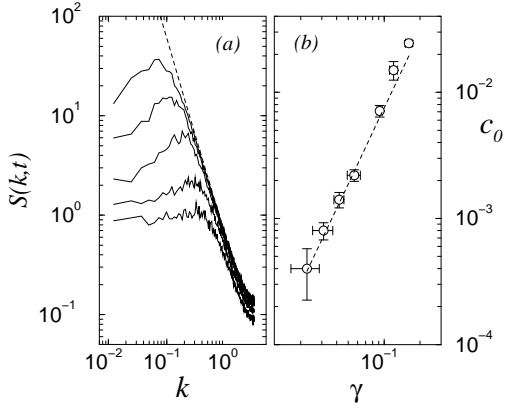, width=2.2in, angle=-90}
\caption{(a) Structure factor of the PE for
$t=1000$, $5248$, $29512$, $165960$, and $10^6$ MC units (bottom to top). 
Each solid line is an average over 200 MC realizations with 
$A=10$, $J=1$, $k_B T = 1/3$, and $L_x=512$. Units are arbitrary. 
Dashed line is the exact asymptotic structure factor 
for the (discretized) Eq.\ (\ref{ec_asymp}), behaving as 
$S(k, \infty) \sim k^{-2}$ for small $k$.
(b) Concentration of holes near the PE as a function
of $\gamma$. Dashed line is the analytical 
prediction (\ref{gamma}) and the circles are results of MC simulations
from $k_B T = 0.27$ up to $k_B T=0.5$, left to right. 
Other parameters are as in panel (a).}
\label{sdek}
\vspace{-0.5cm}
\end{center}
\end{figure}

To study the dynamics of the PE, definitions of spin-percolative
type are needed.  Thus, the {\em precursor film} is the particle cluster
$\Pcal(t)$ in $z=1$ which is connected along nearest neighbor bonds to
the boundary line $y=1$. The PE is at $y=h(x,t)$ for
$x=1,\ldots,L_x$, where $h(x,t)$ is given by the maximum value of $y$
among those cells $(x,y)$ that belong to $\Pcal(t)$.  In Fig.\
\ref{w2} we present results for the dynamics of the PE:
the velocity $V(t) = d\bar s/dt$ [panel (a)], where $\bar s(t)$ is the the
mean PE displacement $\bar s(t) = (1/L_x)
\sum_{x=1}^{L_x} h(x,t)$, and the roughness $w^2(L_x,t) = 
(1/L_x)\sum_{x=1}^{L_x}
\mean{\zeta^2(x,t)}$ [panel (b)], where $\zeta(x,t) \equiv h(x,t) - \bar s(t)$
and $\mean{\cdots}$ stands for average over different MC
runs. 
After a short initial transient $t=\tau_0 \simeq 10^3$ MC units
(independent of $L_x$), we recover the law $\bar \bs(t) = \gamma
t^{1/2}$, with $\gamma$ a time independent constant.  This is in clear
qualitative agreement with the experimental results, and the
confirmation of the universal $\sqrt{t}$ law is considerably more
precise than in the 3-d simulations
\cite{douglas_ising}.  On the other hand, for times longer than the initial
transient $t > \tau_0$ ---before which the roughness oscillates due to
the discreteness of the model and the flat initial condition on the
PE---, the PE displays kinetic roughening as reflected in the behavior
$w(L_x,t) \sim t^{\beta}$ \cite{laszlo}.  The growth exponent $\beta$
seems to cross over from the value $\beta =1/6$ at intermediate times
to $\beta = 1/8$ at long times. Due to the restricted range of
computationally accessible values of $L_x$, the roughness exponent
$\alpha$ in the stationary state relation $w(L_x,\infty) \sim
L_x^{\alpha}$ \cite{laszlo} is more accurately extracted from the
behavior of the structure factor $S(\bk,t) = \mean{{\zeta}_k(t)
\zeta_{-k}(t)}$, where $\zeta_k(t)$ is the Fourier
transform of $\zeta(x,t)$. For a rough interface, $S(\bk,t)$ is
expected \cite{laszlo} to behave as $S(\bk,t) \sim k^{-(2\alpha +1)}$
for long enough times. In Fig.\ \ref{sdek}(a) we show the structure
factor of the PE measured for various times. As is clear
from the figure, $S(\bk,t)$ tends to a scale invariant behavior at
long times, characterized by a roughness exponent $\alpha = 1/2$.

Inspection of typical runs shows a rather sparse distribution of
particles [concentration $c_\rp(\br,t)$] and holes [concentration
$c_\rh(\br,t)$], which attenuates very rapidly away from their respective
sources\cite{no_freespace}. Since the concentrations are
typically low, the dynamics suggest to introduce 
an independent and non-interacting particle-hole diffusion
model, with identical equations for particles and holes.
We will treat the $\br$ coordinate as a
continuous variable. Thus, on $\Pcal(t)$, we have
\beq\label{diffusion}
\partial_t c_\rh= D \Delta c_\rh - \nabla \cdot \bq,
\eeq
where $\bq(\br,t)$ is a noise term representing fluctuations in the hole
diffusion current within $\Pcal(t)$ \cite{karma_misbah}. 
The interface velocity along the outward normal $\bn$ is given by
\beq\label{bc_1}
V_n = 2 (D \nablab c_\rh -\bq) \cdot \bn - B \nabla_s^2 \kappa
- \nabla_s \cdot \bp.
\eeq
The first term on the right in (\ref{bc_1}) comes from diffusion of
holes away from the PE. The prefactor two reflects the dual r\^ole
of particles and holes and that both have the same diffusion
constant. The second term represents {\em surface diffusion} due to the
tendency for particles/holes to move along the interface from regions
of negative to positive curvature, the constant $B$ being proportional
to the interface concentration of mobile species. Here $\kappa$ is the
mean curvature.  Finally, $\bp(\br,t)$ is a noise term describing
fluctuations in the interface diffusion current along the precursor
edge \cite{tang_nattermann}. The concentration obeys also the boundary
condition $c_\rh(x,0,t)=0$ and the Gibbs-Thompson condition \cite{abs}
\beq\label{bc_2}
\left. c_\rh(\br,t)\right|_{y=h(x,t)} = c_0 + \Gamma \kappa,
\eeq
where $c_0$ is the average concentration of holes at the precursor
edge and $\Gamma = c_0\sigma/k_B T$, where $\sigma$ is the surface 
tension, which to leading order is taken to be isotropic. 

This moving boundary problem presents formidable obstacles to direct
attack. Neglecting the noise terms $\bq$ and $\bp$, assuming a flat
interface, and averaging over $x$, a classical Stefan problem \cite{crank2} is
recovered for the mean hole concentration $\bar c(y,t)$. On $\Pcal(t)$, 
$\bar c(y,t)$ satisfies the diffusion equation $\partial_t \bar c
= D \partial_y^2 \bar c$, but now with far simpler boundary conditions 
$\left. \bar c(y,t)\, \right|_{y=\bar s(t)}=c_0$, 
$\bar c(0,t)=0$ and 
\beq\label{bc_1medio}
\frac{d \bar s}{dt} = 2 D \left. \partial_y \bar c\, \right|_{y=\bar s(t)}.
\eeq
This problem admits a similarity solution \cite{crank2} in which
$\bar s(t) = \gamma t^{1/2}$, and
\beq\label{conc_0}
\bar c(y,t) = \frac{\gamma}{4D} \int_0^{y/t^{1/2}} e^{(\gamma^2-u^2)/4D}\ du,
\eeq
where $\gamma$ is the unique solution of
\beq\label{gamma}
c_0 = \frac{\gamma}{4D} \int_0^\gamma e^{(\gamma^2-u^2)/4D}\ du. 
\eeq
Equation (\ref{gamma}) then gives a relation between $\gamma$ and
$c_0$. Now, both $\gamma$ and $c_0$ can be read off from the
simulations, with different choices of temperature giving different
$(c_0,\gamma)$ points which are compared in Fig.\ \ref{sdek}(b) with the
predictions of (\ref{gamma}). The way the simulations of the lattice
gas are set up implies that, for comparison with the diffusive system,
we should have $D=1/3$.  The excellent agreement in Fig.\ \ref{sdek}(b)
furnishes very significant support for our continuum model.
 
The spatio-temporal fluctuations of the PE, expressed by $w(L_x,t)$
and $S(q,t)$ in Figs.\ \ref{w2}(b) and \ref{sdek}(a), respectively, require
full implementation of the boundary conditions (\ref{bc_1}) and
(\ref{bc_2}), a formidable mathematical problem.  Progress can be made
when $\gamma$ is small [see Fig.\ \ref{sdek}(b)] by 
{\em (i)} setting $\partial_t c_\rh = 0$ for $y < h(x,t)$, the
quasistationary approximation \cite{langer}; {\em (ii)} perturbing around the 
spatially-averaged
solution given by (\ref{conc_0}) and (\ref{gamma}) \cite{doole}:
we expand Eq.\ (\ref{bc_1}) to quadratic order in the fluctuation variable 
$\zeta(x,t) = h(x,t) - \bar s(t)$ describing the precursor edge
---taking the noise terms to be of the same order as $\zeta(x,t)$
\cite{karma_misbah}---, to obtain 
\begin{eqnarray}
\frac{\partial \zeta_k}{\partial t}&= &- \left[D |k| \left(\Gamma k^2 + 
\frac{c_0}{\gamma 
t^{1/2}}\right) \coth(|k| \gamma t^{1/2})+B k^4 \right]\,\zeta_k \nonumber \\
& & +
\frac{V(t)}{2}\Fcal_k[(\partial_x \zeta)^2] +  \eta_k(t)\label{kpz_like},
\end{eqnarray}
where $\Fcal_k[f(x)]$ is the Fourier transform of $f(x)$, and $\eta_k(t)$
is a white noise with correlations \cite{noises}
\begin{eqnarray}
\mean{\eta_k(t) \eta_{k'}(t')}&=& 2 c_0 \delta_{k,-k'} \delta(t-t') 
\nonumber \\
&\times&\left\{D \left[1+\frac{|k|^2 \gamma
t^{1/2}}{\sinh^2(|k|\gamma t^{1/2})}\right] + \frac{B}{\Gamma} k^2\right\}.
\label{noise_corr}
\end{eqnarray}
In Eq.\ (\ref{kpz_like}) the linear terms stabilize PE fluctuations, 
while the nonlinear term is of the Kardar-Parisi-Zhang (KPZ) type
\cite{kpz}, with a coefficient $V(t) = \gamma/t^{1/2}$ that decays in time. 
In the asymptotic
regime $t \to \infty$, and neglecting terms proportional to $c_0$, 
$\Gamma \sim \Ocal(\gamma^2)$, Eqs.\ (\ref{kpz_like}) and 
(\ref{noise_corr}) reduce to the conserved linear equation
\beq \label{ec_asymp}
\frac{\partial \zeta_k}{\partial t} = -B k^4 \zeta_k + \eta_k,
\eeq
with noise correlations $\mean{\eta_k(t) \eta_{k'}(t')} = 2 (c_0
B/\Gamma) k^2 \delta_{k,-k'} \\ \delta(t-t')$. It is easy to solve
this equation exactly and obtain $\beta= 1/8$ and $\alpha=1/2$, which
provide accurate fits at long times for the behavior of the roughness
[Fig.\ \ref{w2}(b)] and the structure factor [Fig.\ \ref{sdek}(a)],
respectively.  Moreover, the behavior of $S(k,t)$ at {\em long} but
{\em finite} times [$e.g.$, $t=10^6$ MC units in Fig.\ \ref{sdek}(a)]
is also consistent with that predicted by Eq.\
(\ref{ec_asymp}). 
For sufficiently long times, the rate of arrival of particles becomes
essentially zero. In this case, emission of particles to the exterior of 
the PE and holes to the interior are in balance, so the mean interface
is stationary.
In such a
situation, up to order $\Ocal (\gamma^2)$, the only fluctuations on
the PE are due to the conserved surface diffusion current, which is
precisely described by Eq.\ (\ref{ec_asymp}).
However, for intermediate times ($e.g.$, $t \lesssim 10^5$ MC units), 
the behavior of the fluctuations deviates from that predicted by 
Eq.\ (\ref{ec_asymp}), see Figs.\ \ref{w2}(b) and \ref{sdek}(a).
A plausible explanation
is that the KPZ term in Eq.\ (\ref{kpz_like}) is still non-negligible, 
in which case it should dominate the large scale properties of the PE.
Using results for the KPZ equation \cite{laszlo,hgang}, the roughness 
should then be given in this time regime by $w \sim [V(t) \, t]^{1/3} \sim 
t^{1/6}$, again in rather good agreement with the numerical results
for the lattice gas, see Fig.\ \ref{w2}(b).

In summary, using simulations of our new driven lattice gas model we
have {\em (i)} recaptured the $\sqrt{t}$ universal mean precursor
displacement law; {\em (ii)} shown that the spatio-temporal
fluctuations of the PE demonstrate dynamical scaling; 
{\em (iii)} shown that the
mechanisms of matter transport from the reservoir droplet to the PE
involve low density of particles and holes, described by diffusion
equations with moving boundaries; {\em (iv)} derived a stochastic
partial differential equation for the PE which brings in the
fluctuation regimes found previously in a natural way, and which tends
asymptotically to a Langevin description for a free interface with 
conserved dynamics as in
(\ref{ec_asymp}), possibly observable experimentally, even though it
presents a KPZ behavior at intermediate times.
In preliminary work, we have investigated our lattice gas model for a
single layer ($z=1$) in a thermal gradient, finding a mean spatial 
displacement growing linearly in time, generated by biased hole diffusion. 
This has relevance for microfluidics \cite{troian}. 

We thank A. S\'anchez for participation in the early
stages of this work, and
M. Castro for discussions. E.M.\ acknowledges the EU fellowship
No.\ HPMF-CT-2000-0487. This research has been supported by EPSRC (UK) Grant
No.\ GR/M04426 and by DGES (Spain) grants Nos.\ HB1999-0018 and BFM2000-0006.


\begin{thebibliography}{99}
\bibitem{no_freespace} There is also emission of particles from the precursor
edge into the free space ahead of it. This gas of particles is close 
to equilibrium with the interface and thus, its effect will be neglected in 
Eq. (\ref{bc_1}).
\bibitem{abs} In Eq. (\ref{bc_2}) fluctuations \cite{karma_misbah}
have been suppressed due to the absorbing boundary condition implicit in Eq.
(\ref{bc_1}).
\bibitem{noises} The correlations for $\eta_k(t)$ originate from those
for $\bq$ in Eq. (\ref{diffusion}) and $\bp$ in Eq. (\ref{bc_1}). A local
equilibrium hypothesis determines these to be $\mean{q_i(\br,t) q_j(\br',t')}
= 2 D c_0 \delta_{ij}\delta(\br-\br')\delta(t-t')$ and $\mean{p_i(\br,t)
p_j(\br',t')}= 2 (B c_0/ \Gamma) \delta_{ij}\delta(\br-\br')\delta(t-t')$,
see \cite{karma_misbah,tang_nattermann}.
\end{thebibliography}
\end{document}